\documentstyle{npbproc}
% ...
\newcommand{\be}{\begin{equation}}
\newcommand{\ee}{\end{equation}}
\newcommand{\<}{\langle}
\renewcommand{\>}{\rangle}
\def\reff#1{(\protect\ref{#1})}
\newcommand{\C}{{\cal C}}
\def\spose#1{\hbox to 0pt{#1\hss}}
\def\ltapprox{\mathrel{\spose{\lower 3pt\hbox{$\mathchar"218$}}
 \raise 2.0pt\hbox{$\mathchar"13C$}}}
\def\gtapprox{\mathrel{\spose{\lower 3pt\hbox{$\mathchar"218$}}
 \raise 2.0pt\hbox{$\mathchar"13E$}}}
\def\br{{\bf r}}

\begin{document}
\title{DYNAMIC CRITICAL BEHAVIOUR OF WOLFF'S ALGORITHM FOR
  $RP^N$ $\sigma$-MODELS}
\author{Sergio Caracciolo\vspace*{-3mm}}
\address{Scuola Normale Superiore, Pisa, Italy\vspace*{-2mm}}
\author{Robert G. Edwards\vspace*{-3mm}}
\address{SCRI,  Florida State University, Tallahassee, USA\vspace*{-2mm}}
\author{Andrea Pelissetto\vspace*{-3mm}}
\address{Dipartimento di Fisica, Universit\`a di Pisa, Italy\vspace*{-2mm}}
\author{Alan D. Sokal\vspace*{-3mm}}
\address{Department of Physics, New York University, USA}

\date{}

\runtitle{Wolff-type algorithms for $RP^N$ $\sigma$-models}
\runauthor{S. Caracciolo et al.}
\volume{XXX}  
\firstpage{1} 
\lastpage{3}  

\begin{abstract}
We study the performance of a Wolff-type embedding algorithm for $RP^N$
$\sigma$-models.  We find that the
algorithm in which we update the embedded Ising model \`a la  Swendsen-Wang
has critical slowing-down as $z_\chi \approx 1$.
If instead we update the Ising spins with a perfect algorithm which at every
iteration produces a new independent configuration,
we obtain  $z_\chi \approx 0$.
This shows that the Ising embedding encodes well the collective modes of the
system, and that the behaviour of the first algorithm is connected to the poor
performance of the Swendsen-Wang algorithm in dealing with a frustrated Ising
model.
\end{abstract}

\maketitle


In recent years there has been a lot of work in devising new algorithms which,
 by taking into proper account the collective modes of the theory, are able to
eliminate or at least to reduce critical slowing-down.

For $O(N)$ $\sigma$-models, an extremely efficient algorithm was proposed three
years ago by Wolff \cite{Wolff}.
In two dimensions, numerical tests of the dynamic critical
behaviour show the complete or almost complete absence of critical slowing-down
(i.e. $z \ltapprox 0.1$) \cite{Wolff,ES}.

The extraordinary efficiency of this algorithm has spurred many attempts to
find
generalizations to $\sigma$-models taking values in manifolds other than
spheres. However, last year \cite{CEPS-Tallahassee} we presented a heuristic
argument whose conclusion was:
{\sl a necessary condition for a Wolff-type embedding algorithm
to work well}
(even with {\sl perfect} updating of the induced Ising spins)
{\sl is that the manifold is a sphere, a real projective space, or a discrete
quotient of products of such spaces} \cite{CEPS}.

Let us briefly review the general principles of Wolff-type embedding
algorithms \cite{CEPS,Sokal_LAT90}. Consider a general
$\sigma$-model taking values in a Riemanian manifold $M$, with
Hamiltonian of the form
\be
H(\{\sigma\}) \,=\, \beta \sum_{\< xy \>}\, E(\sigma_x,\sigma_{y}).
\ee
Then the algorithm is defined by a collection of energy-preserving maps $T$,
and gives rise to the induced Ising Hamiltonian
\begin{eqnarray}
H(\{\epsilon\})   & = &   - \sum_{\< xy\>}\, J_{xy} \epsilon_x \epsilon_{y}
                                                                \nonumber  \\
   &  &  \quad   - \sum_{\< xy\>}\, h_{xy} ( \epsilon_x -
\epsilon_{y})\, +\, \hbox{\rm const}
  \label{eqn2}
\end{eqnarray}
where $\{\epsilon\}$ are Ising spins and
\begin{eqnarray}
J_{xy} & = & {\beta\over4}\, [E(T\sigma_x,\sigma_y)\,
+\,E(\sigma_x,T\sigma_y)\,   -\, 2 E(\sigma_x,\sigma_y)]   \nonumber\\
h_{xy} & = & {\beta\over4}\, [E(T\sigma_x,\sigma_y)\,
-\,E(\sigma_x,T\sigma_y)]
\end{eqnarray}
In practice an iteration of the algorithm works as follows:
\begin{itemize}
\begin{enumerate}
\item  Choose a map $T$ in the given family according to a given
distribution.
\item Initialize all Ising spins $\epsilon_x = 1$.
\item Update the embedded Ising model.
\item Set $\sigma_x = T\sigma_x$ where $\epsilon_x = -1$.
\end{enumerate}\end{itemize}

In step (iii) one can use any valid algorithm for simulating the Ising model
\reff{eqn2}. We will consider two different choices:

a) The {\em practical}\/ algorithm where step (iii) consists of one
standard (full-lattice) Swendsen-Wang update.

b) The {\em idealized}\/ algorithm where at step (iii) we generate a new
configuration of Ising spins, independent of the old one. This is achieved in
practice by performing at every iteration  $N_{hit}$ Swendsen-Wang updates,
where $N_{hit}$ is chosen so large that the autocorrelation times of the
various observables are independent of $N_{hit}$ within error bars.

The idealized algorithm allows us to understand how well the embedding
succeeds in embodying the important large-scale collective modes of the
$\sigma$-model.
A bad performance of the idealized algorithm means that in the $\sigma$-model
there are other important excitations which are not captured by the embedding.
By contrast, a poor performance of the practical algorithm might be due solely
to the bad performance of the algorithm used in updating the Ising spins.

What we have defined is a generalization of Wolff's algorithm for $O(N)$
models,
and we claim \cite{CEPS} that it can work well only in a few cases.
The reason for this is that in order to perform well the algorithm must do a
good job in handling the collective modes of the theory,
which certainly include long-wavelength spin-waves.
In order to treat these modes well, we argue that
the set of links for which $J_{ij} \approx 0$ must disconnect the $x$-space
into
two or more regions. It follows \cite{CEPS} that the embedding map must have
the
{\em codimension-1 property}\/:
the fixed-point manifold of the map $T$ must have codimension 1.
Differential geometry can then be used to prove that the only
manifolds which satisfy this requirement are $S^N$ or $RP^N$ (and discrete
quotients of products thereof).

Let us notice that our heuristic argument gives a {\em necessary}\/
condition for the idealized (and hence also the practical)
algorithm to beat critical slowing-down, but it does not guarantee that either
the idealized or the practical algorithm will in fact perform well.
For this reason we have decided to study
the two-dimensional $RP^N$ model.

The real projective space $RP^{N-1}$ is by definition the sphere $S^{N-1}$ with
antipodal points identified, i.e. $RP^{N-1} = S^{N-1}/Z_2$.
The most convenient approach is to consider spins taking
values on the sphere $S^{N-1}$,
subject to the condition that the Hamiltonian and
all physical observables must be invariant under the $Z_2$ local gauge
transformations $\sigma_x \to \eta_x\sigma_x$ with $\eta_x = \pm 1$.  The
simplest lattice Hamiltonian for this model is therefore
\be
H(\{\sigma\}) \,=\, -{\beta\over2} \sum_{x,\mu} \,
   (\sigma_x\cdot\sigma_{x+\mu})^2
\ee
The continuum limit of this model is not at all clear.
In the formal continuum limit $a\to 0$,
the Hamiltonian becomes that of the continuum $O(N)$ non-linear $\sigma$-model.
In order to explain why in the continuum limit the theory does
not have the $Z_2$ gauge invariance, it has been suggested
\cite{Caselle-Gliozzi} that at a finite value of the coupling
the system undergoes a
phase transition which gives rise to a condensation of the vortices.
However, the presence of this phase transition is rather controversial
(see \cite{Caselle-Gliozzi,Sinclair,Chiccoli,Kunz} and references therein).
We do not have yet much to add to this point,
and in the following we will address the problem of
the dynamical behaviour of the algorithm.

The algorithm is defined by the same embedding used by Wolff for $O(N)$
$\sigma$-models:
the induced Hamiltonian is given by (2) with $h_{xy}=0$ and
\be
J_{xy} \, = \,  \beta (\sigma^\bot_x \cdot \sigma^\bot_y)
                               (\sigma_x \cdot \br)  (\sigma_y \cdot \br)  \;,
\ee
where $\sigma^\bot_x = \sigma_x - (\sigma\cdot \br) \br$.
Let us notice that, when $N \ge 3$, the induced Hamiltonian is frustrated.

Let us first discuss the behaviour of the practical algorithm. We have measured
the energy, the tensor susceptibility $\chi_T$ and the correlation length in
the
tensor channel $\xi$ for both $RP^2$ and $RP^3$ on lattices of dimension
$L=32,64,128$
(a detailed discussion of the simulation is given in \cite{CEPS_RP}).

A finite-size scaling analysis of $L^{-z_\chi} \tau_{int,\chi}$ versus $\xi/L$
shows that the points are well fitted using
\be
z_{int,\chi} = \cases{ 0.9\pm 0.3 & for $RP^2$ \cr 1.1\pm 0.3 & for $RP^3$\cr}
\ee
while a similar analysis for the energy gives
\be
z_{int,E} = \cases{ 0.2\pm 0.3 & for $RP^2$ \cr 0.2\pm 0.3 & for $RP^3$\cr}
\ee
This means that the practical algorithm,
though providing a significant improvement over local algorithms,
still suffers from strong critical slowing-down.
At this point, however, it is not clear what is the cause of this behavior:
are there other excitations in the model which are not well encoded in the
embedding, or is the critical slowing-down due instead to the Swendsen-Wang
subroutine which is unable to simulate efficiently a frustrated Ising model?

To answer this question we have studied the idealized algorithm for $RP^2$
on lattices with $L=32,64$. We have found that the dynamic critical exponent
for the susceptibility is now
\be
z_{int,\chi} \, =\, 0.1 \pm 0.3
\ee
Critical slowing-down is thus nearly eliminated!
We conclude that the embedding encodes well
the collective modes of the $RP^N$ model,
and that the failure of the practical version must be
ascribed to the Ising subroutine.

We thank Paolo Pasini and Ulli Wolff  for helpful
discussions and for sending us their unpublished data.
The computations reported here were performed at the Centro di Calcolo, SNS,
Pisa. The authors' research was supported in part by
INFN, U.S. DOE contracts DE-FC05-85ER\-250000 and DE-FG02-90\-ER\-40581,
NSF grant DMS-8911273, and NATO CRG-910251.


\end{document}